\documentstyle[amssymb,epsfig]{fbssuppl}
\begin{document}
\title{Resonant-state solution of the
Faddeev-Merkuriev integral equations for three-body systems with Coulomb-like
potentials}
\author{Z.\ Papp${}^{1,2}$,  S.\ L.\ Yakovlev${}^{3}$, C-.Y.\ Hu${}^{1}$,  
J.~Darai${}^{4}$, I.\ N.\ Filikhin${}^{3}$ and B.~K\'onya${}^{2}$}
\institute{${}^{1}$ Department of Physics and Astronomy, 
California State University, Long Beach, CA 90840, USA \\
${}^{2}$ Institute of Nuclear Research of the
Hungarian Academy of Sciences, 
Bem t\'er 18/C, H--4026 Debrecen, Hungary \\
${}^{3}$ Department of Mathematical and Computational Physics, 
St.\ Petersburg State University,
198904 St.\ Petersburg, Petrodvoretz,
Ulyanovskaya Str. 1, Russia \\
${}^{4}$ Department of Experimental Physics, University of Debrecen, 
Bem t\'er 18/A, H--4026 Debrecen, Hungary 
}
\date{\today}
\maketitle

\begin{abstract}
\noindent
A novel method for calculating resonances in three-body
Coulombic systems is presented. The Faddeev-Merkuriev integral equations
are solved by applying the Coulomb-Sturmian separable expansion method.
To show the power of the method we calculate resonances
of the three-$\alpha$ and the $H^-$ systems.
\end{abstract}
%\vspace{0.5cm}

\section{Introduction}

For three-body systems the Faddeev equations are the fundamental equations.
After one iteration they possess connected kernels,  consequently they are
effectively Fredholm integral equations of second kind. Thus the Fredholm
alternative applies: at certain energy either the homogeneous or the 
inhomogeneous
equations have solutions. Three-body bound states correspond to
the solutions of the homogeneous Faddeev equations at real energies,
resonances, as usual in quantum mechanics,
are related to their complex-energy solutions.

The Faddeev equations were derived for short-range
interactions and if we simply plug-in a Coulomb-like potential they become
singular. The necessary modifications have been formulated
 in the Faddeev-Merkuriev theory \cite{fm-book}
on a mathematically sound and elegant way via
integral equations with connected (compact) kernels and
configuration space differential equations 
with asymptotic boundary conditions.

Recently, one of us has developed a novel method 
for treating the three-body Coulomb problem
via solving the set of Faddeev-Noble and Lippmann-Schwinger
integral equations in Coulomb--Sturmian-space representation.
The method was elaborated first
for bound-state problems \cite{pzwp} with repulsive Coulomb plus nuclear
potential, then it was extended for calculating $p-d$
scattering at energies below the breakup threshold \cite{pzsc}.
Also atomic  bound-state problems with attractive Coulomb
interactions were considered \cite{pzatom}.
In these calculations an excellent agreement with the results of other well
established methods were found and
the efficiency and the accuracy of the method were demonstrated.

More recently we have extended the method  for calculating 
resonances in three-body systems with short-range plus repulsive
Coulomb interactions by solving the Faddeev-Noble integral
equations \cite{zis}. 
Here we solve the Faddeev-Merkuriev integral equations. This way we can
handle all kind of Coulomb-like potentials, not only repulsive but also
attractive ones. For illustrating the power  of this method
we show our previous three-$\alpha$ results and, as novel feature,
we calculate resonances in  the $H^-$ ($pee$) system. 

\section{Faddeev-Merkuriev integral equations}
The Hamiltonian of a three-body Coulombic  system reads
\begin{equation}
H=H^0 +W + v_\alpha+ v_\beta + v_\gamma,
\label{H}
\end{equation}
where $H^0$ is the three-body kinetic energy
operator, $W$ stands for the possible
three-body potential and $v_\alpha$ denotes the
Coulomb-like interaction in the subsystem $\alpha$.
We use throughout the usual
configuration-space Jacobi coordinates
$x_\alpha$  and $y_\alpha$.
Thus  $v_\alpha$ only depends on $x_\alpha$ ($v_\alpha=v_\alpha (x_\alpha)$),
while $W$ depends on both  $x_\alpha$ and $y_\alpha$ coordinates
($W=W(x_\alpha,y_\alpha)$).

The physical role of a Coulomb-like potential is twofold.
Its long-distance part modifies the asymptotic
motion, while its short-range part strongly correlates the
two-body subsystems.
Merkuriev proposed to split the potentials into short-range and long-range
parts in the three-body configuration
space via a cut-off function $\zeta$,
\begin{equation}
v_\alpha^{(s)}(x_\alpha,y_\alpha)=
v_\alpha(x_\alpha) \zeta_\alpha(x_\alpha,y_\alpha),\end{equation}
and
\begin{equation}v_\alpha^{(l)}(x_\alpha,y_\alpha)=
v_\alpha(x_\alpha) [1- \zeta_\alpha(x_\alpha,y_\alpha) ].
\label{potm}
\end{equation}
The function $\zeta_\alpha$ is defined such that it separates the
asymptotic two-body sector $\Omega_\alpha$ from the rest
of the three-body configuration space.
On the region of $\Omega_\alpha$
 the splitting function $\zeta_\alpha$
asymptotically tends to $1$ and  on the complementary asymptotic region
of the configuration space it tends
to $0$. Rigorously, $\Omega_\alpha$ is defined as a part of the
three-body configuration
space where the condition
%%%%%%%%%%%%%%%%%%%%%
\begin{equation}
|x_\alpha| < x_0 (1+|y_\alpha|/y_0)^{1/\nu}, 
\mbox{with} \ \ \ x_0>0,\ y_0>0,\ \nu > 2,
\label{oma}
\end{equation}
%%%%%%%%%%%%%%%%%%%%%
is satisfied. So, in $\Omega_\alpha$ the short-range part
$v_\alpha^{(s)}$ coincides with the
original  Coulomb-like potential $v_\alpha$
and in the complementary region vanishes, whereas
the opposite holds true for $v_\alpha^{(l)}$. 
Note that for repulsive Coulomb interactions
one can also adopt Noble's approach \cite{noble}, where the splitting is
performed in the two-body configuration space. 
This approach can be considered as the $y_0\to\infty$ limit
of Merkuriev's splitting.
Then $v_\alpha^{(l)}$ coincides with the whole Coulomb interaction and
$v_\alpha^{(s)}$ with the short-range nuclear potential.

In the Faddeev procedure we split the wave function into
three components
\begin{equation}
|\Psi \rangle = |\psi_{\alpha} \rangle +
|\psi_{\beta} \rangle +|\psi_{\alpha} \rangle,
\end{equation}
where the components are defined by
\begin{equation}
|\psi_\alpha \rangle = G^{(l)} (z) v_\alpha^{(s)} |\Psi \rangle.
\end{equation}
Here $G^{(l)}$ is the resolvent of the long-ranged Hamiltonian
\begin{equation}
H^{(l)} = H^0 + W +  v_\alpha^{(l)}+
v_\beta^{(l)}+ v_\gamma^{(l)},
\label{hl}
\end{equation}
$G^{(l)}(z)=(z-H^{(l)})^{-1}$, and $z$ is the complex energy parameter.
The wave-function components satisfy the homogeneous
Faddeev-Merkuriev integral equations
\begin{equation}
|\psi_{\alpha} \rangle= G_\alpha^{(l)} (z)
v^{(s)}_\alpha \sum_{\gamma\neq\alpha}
|\psi_{\gamma} \rangle,
\label{fn-eq}
\end{equation}
for $\alpha=1,2,3$, where $G^{(l)}_\alpha$ is the resolvent of the channel 
long-ranged Hamiltonian
\begin{equation}
H^{(l)}_\alpha = H^{(l)} + v_\alpha^{(s)},
\label{hla}
\end{equation}
$G^{(l)}_\alpha(z)=(z-H^{(l)}_\alpha)^{-1}$.
Merkuriev has proved that Eqs.\ (\ref{fn-eq}) possess compact kernels 
for positive $E$ energies, and this property remains valid also for
complex energies $z=E-i\Gamma/2$, $\Gamma > 0$.

\section{Solution method}

We solve these integral equations
by using the Coulomb--Sturmian separable expansion approach.
The Coulomb-Sturmian (CS) functions are defined by
\begin{equation}
\langle r|n \rangle =\left[ \frac {n!} {(n+2l+1)!} \right]^{1/2}
(2br)^{l+1} \exp(-b r) L_n^{2l+1}(2b r),  \label{basisr}
\end{equation}
with $n$ and $l$ being the radial and
orbital angular momentum quantum numbers, respectively, and $b$ is the size
parameter of the basis.
The CS functions $\{ |n \rangle \}$
form a biorthonormal
discrete basis in the radial two-body Hilbert space; the biorthogonal
partner defined  by $\langle r |\widetilde{n }\rangle=
r^{-1} \langle r |{n}\rangle$. 
Since the three-body Hilbert space is a direct product of two-body
Hilbert spaces an appropriate basis
can be defined as the
angular momentum coupled direct product of the two-body bases 
(the possible other
quantum numbers are implicitly assumed)
\begin{equation}
| n \nu \rangle_\alpha =
 | n  \rangle_\alpha \otimes |
\nu \rangle_\alpha, \ \ \ \ (n,\nu=0,1,2,\ldots).
\label{cs3}
\end{equation}
With this basis the completeness relation
takes the form
\begin{equation}
{\bf 1} =\lim\limits_{N\to\infty} \sum_{n,\nu=0}^N |
 \widetilde{n \nu } \rangle_\alpha \;\mbox{}_\alpha\langle
{n \nu } | =
\lim\limits_{N\to\infty} {\bf 1}^{N}_\alpha.
\end{equation}
Note that in the three-body Hilbert space,
three equivalent bases belonging to fragmentation
$\alpha$, $\beta$ and $\gamma$ are possible.

In Ref.\ \cite{pzwp} a novel approximation scheme has been
proposed to the Faddeev-type integral equations
\begin{equation}
|\psi_{\alpha} \rangle= G_\alpha^{(l)} (z)
{\bf 1}^{N}_\alpha v^{(s)}_\alpha \sum_{\gamma\neq\alpha}
{\bf 1}^{N}_\gamma |\psi_{\gamma} \rangle,
\label{feqsapp}
\end{equation}
i.e.\ the short-range potential
$v_\alpha^{(s)}$ in the three-body
Hilbert space is taken to have a separable form, viz.
\begin{equation}
v_\alpha^{(s)} = 
\lim_{N\to\infty} {\bf 1}^{N}_\alpha v_\alpha^{(s)} {\bf 1}^{N}_\beta 
\approx {\bf 1}^{N}_\alpha v_\alpha^{(s)} {\bf 1}^{N}_\beta
= \sum_{n,\nu ,n^{\prime },
\nu ^{\prime }=0}^N|\widetilde{n\nu }\rangle _\alpha \;
\underline{v}_{\alpha \beta }^{(s)}
\;\mbox{}_\beta \langle \widetilde{n^{\prime }
\nu ^{\prime } }|,  \label{sepfe}
\end{equation}
where $\underline{v}_{\alpha \beta}^{(s)}=
\mbox{}_\alpha \langle n\nu |
v_\alpha^{(s)}|n^{\prime }\nu ^{\prime} \rangle_\beta$.
The validity of the approximation relies on 
the square integrable
property of the term $v_\alpha^{(s)} |\psi_\gamma \rangle$, 
$\gamma \neq \alpha$. Thus this approximation is
justified also for complex energies as long as this property remains valid.
In Eq.~(\ref{sepfe}) the ket and bra states are defined
for different fragmentation, depending on the
environment of the potential operators in the equations.
Now, with this approximation, the solution of the homogeneous
Faddeev-Merkuriev equations
turns into solution of matrix equations for the component vector
$\underline{\psi}_{\alpha}=
 \mbox{}_\alpha \langle \widetilde{ n\nu } | \psi_\alpha  \rangle$
\begin{equation}
\underline{\psi}_{\alpha} = \underline{G}_\alpha^{(l)} (z)
\underline{v}^{(s)}_\alpha \sum_{\gamma\neq\alpha}
\underline{\psi}_{\gamma},
\label{feqm}
\end{equation}
where $\underline{G}_\alpha^{(l)}=\mbox{}_\alpha \langle \widetilde{
n\nu } |G_\alpha^{(l)}|\widetilde{n^{\prime}\nu^{\prime}
 }\rangle_\alpha$. A unique solution exists if
and only if
\begin{equation}
\det \{ [ \underline{G}^{(l)}(z)]^{-1} - \underline{v}^{(s)} \} =0.
\end{equation}

The Green's operator ${G}_\alpha^{(l)}$ is a solution of
the auxiliary three-body problem with the Hamiltonian ${H}_\alpha^{(l)}$.
To determine it uniquely
one should start again from Faddeev-type integral equations, which does not
seem to lead any further, or from the triad of
Lippmann-Schwinger equations \cite{glockle}.
The triad of Lippmann-Schwinger equations, although they do not possess
compact kernels, also define the solution in an unique
way. They are, in fact, related to the adjoint representation
of the Faddeev operator \cite{sl}.
The Hamiltonian $H_\alpha^{(l)}$, however, has a peculiar
property that it supports bound state only in the subsystem $\alpha$, and
thus it has only one kind of asymptotic channel, the $\alpha$ channel.
For such a system one single Lippmann-Schwinger equation is sufficient for
an unique solution \cite{sandhas}.

The appropriate equation takes the form
\begin{equation}
G_\alpha^{(l)}=\widetilde{G}_\alpha +
\widetilde{G}_\alpha  U^\alpha G_\alpha^{(l)},
\label{lsgc}
\end{equation}
where $\widetilde{G}_\alpha$ is the resolvent of the 
channel-distorted long-range Hamiltonian,
\begin{equation}
\widetilde{H}_\alpha=H^0+v_\alpha+u_\alpha^{(l)},
\label{htilde}
\end{equation}
and  $U^\alpha=W + v_\beta^{(l)}+v_\gamma^{(l)}  -u_\alpha^{(l)}$.
The auxiliary potential
$u_\alpha^{(l)}$ depends on the coordinate $y_\alpha$
and  has the asymptotic form
$u_\alpha^{(l)} \sim {e_\alpha (e_\beta+e_\gamma) }/{y_\alpha}$
as ${y_\alpha \to \infty}$. In fact, $u_\alpha^{(l)}$
is an effective Coulomb interaction between the center of
mass of the subsystem $\alpha$ (with
charge $e_\beta+e_\gamma$) and the third particle
(with charge $e_\alpha$). Its role is to compensate the Coulomb tail of the
potentials $v_\beta^{(l)}+v_\gamma^{(l)}$ in $\Omega_\alpha$.

It is important to realize that in this approach to get the solution only
the matrix elements  $\underline{G}_\alpha^{(l)}$ are needed, i.e.
only the representation of the Green's operator
on a compact subset of the Hilbert space are required.
So, although Eq.~(\ref{lsgc}) does not possess a compact kernel on the whole
three-body Hilbert space its matrix form is effectively a compact equation
on the subspace spanned by finite number of CS functions.
Thus we can perform
an approximation, similar to Eq.\ (\ref{sepfe}),
on the potential $U^\alpha$ in Eq.~(\ref{lsgc}),
with bases of the same fragmentation $\alpha$
applied  on both sides of the operator. Now the integral
equation reduces to an analogous set of linear algebraic
equation with the operators replaced by their matrix representations.
The solution is given by
\begin{equation}
[\underline{G}_\alpha^{(l)}(z)]^{-1} =
[\underline{\widetilde{G}}_\alpha (z)]^{-1} - \underline{U}^{\alpha}.
\end{equation}

The most crucial point in this procedure is the
 calculation of the matrix elements
$\underline{\widetilde{G}}_{\alpha}=
\mbox{}_\alpha \langle \widetilde{n\nu } |
\widetilde{G}_\alpha |  \widetilde{ n^{\prime }\nu^{\prime}
}\rangle_\alpha $, since the  potential
matrix elements $\underline{v}^{(s)}_{\alpha \beta}$ and
$\underline{U}^{\alpha}$ can always be calculated numerically by making use of
the transformation of Jacobi coordinates.
The Green's operator $\widetilde{G}_\alpha$
is a resolvent of the sum of two commuting Hamiltonians,
$\widetilde{H}_\alpha = h_{x_\alpha}+h_{y_\alpha}$,
where $h_{x_\alpha}=h^0_{x_\alpha}+v_\alpha$ and
$h_{y_\alpha}=h^0_{y_\alpha}+u_\alpha^{(l)}$,
which act in different two-body Hilbert spaces.
Thus, using  the convolution theorem the three-body Green's operator
$\widetilde{G}_\alpha$ equates to
a convolution integral of two-body Green's operators, i.e.
\begin{equation}
\widetilde{G}_\alpha (z)=
 \frac 1{2\pi \mathrm{i}}\oint_C
dz^\prime \,g_{x_\alpha }(z-z^\prime)\;
g_{y_\alpha}(z^\prime),
 \label{contourint}
\end{equation}
where
$g_{x_\alpha}(z)=(z-h_{x_\alpha})^{-1}$  and
$g_{y_\alpha}(z)=(z-h_{y_\alpha})^{-1}$.
The contour $C$ should be taken  counterclockwise
around the continuous spectrum of $h_{y_\alpha }$
so that $g_{x_\alpha }$ is analytic in the domain encircled
by $C$.

\begin{figure}
\psfig{file=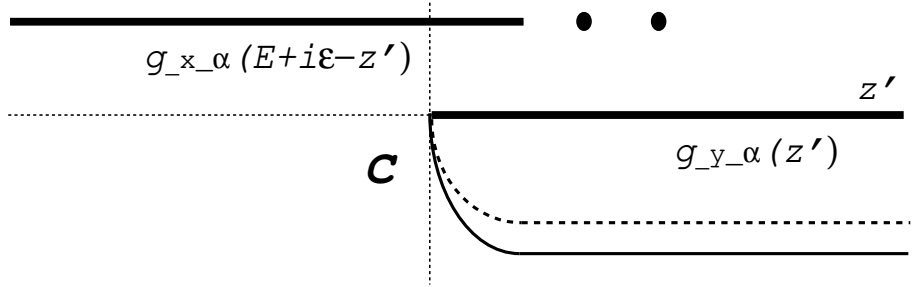,width=12cm}

\caption{Analytic structure of $g_{x_\alpha }(z-z^\prime)\;
g_{y_\alpha}(z^\prime)$ as a function of $z^\prime$ with
$z=E+{\mathrm{i}}\varepsilon$, $E>0$, $\varepsilon>0$.
The contour $C$ encircles the continuous spectrum of
$h_{y_\alpha}$. A part of it, which goes on the unphysical
Riemann-sheet of $g_{y_\alpha}$, is drawn by broken line.}

\label{fig1}
\end{figure}

To examine the structure of the integrand let us
shift the spectrum of $g_{x_\alpha }$ by
taking  $z=E +{\mathrm{i}}\varepsilon$  with
positive $\varepsilon$. By doing so,
the two spectra become well separated and
the spectrum of $g_{y_\alpha}$ can be encircled.
Next the contour $C$ is deformed analytically
in such a way that the upper part descends to the unphysical
Riemann sheet of $g_{y_\alpha}$, while
the lower part of $C$ can be detoured away from the cut
 [see  Fig.~\ref{fig1}]. The contour still
encircles the branch cut singularity of $g_{y_\alpha}$,
but in the  $\varepsilon\to 0$ limit it now
avoids the singularities of $g_{x_\alpha}$.
Moreover, by continuing to negative values of  $\varepsilon$, in order that
we can calculate resonances, the branch cut and pole singularities of
$g_{x_\alpha}$ move
onto the second Riemann sheet of $g_{y_\alpha}$ and, at the same time,
the  branch cut of $g_{y_\alpha}$ moves onto the second Riemann sheet
of $g_{x_\alpha}$. Thus, the mathematical conditions for
the contour integral representation of $\widetilde{G}_\alpha (z)$ in
Eq.~(\ref{contourint}) can be fulfilled also for complex energies
with negative imaginary part.
In this respect there is only a gradual difference between the
bound- and resonant-state calculations. Now,
the matrix elements $\underline{\widetilde{G}}_\alpha$
can be cast in the form
\begin{equation}
\widetilde{\underline{G}}_\alpha (z)=
 \frac 1{2\pi \mathrm{i}}\oint_C
dz^\prime \,\underline{g}_{x_\alpha }(z-z^\prime)\;
\underline{g}_{y_\alpha}(z^\prime),
\label{contourint2}
\end{equation}
where the corresponding CS matrix elements of the two-body Green's operators in
the integrand are known analytically for all complex energies \cite{cpc},
and thus the convolution integral can be performed also in practice.

\section{Numerical illustration}

\subsection{Resonances in a model three-alpha system}

To show the power of this method
we examine the convergence of the results for three-body
resonant-state energies. For this purpose we take first the same
model that has been presented by us before in Ref.\ \cite{zis}.
This is an
Ali--Bodmer-type model for the
charged three-$\alpha$ system interacting via $s$-wave
short-range interaction.
To improve its properties we add a phenomenological
three-body potential.
Adopting Noble's splitting we have
%%%
\begin{equation}
v_\alpha^{(s)}(r)=  V_{1} \exp\{ -r^2/{\beta_1}^2\} +
V_{2} \exp\{ -r^2/{\beta_2}^2\}
\label{filpot}
\end{equation}
with $V_{1}=125$ MeV, $V_{2}=-30.18$ MeV, ${\beta}_1=1.53$ fm,
${\beta}_2=2.85$ fm, and
%%%
\begin{equation}
v_\alpha^{(l)}(r)=  4 e^2/r.
\end{equation}
%%%
We use units such that $\hbar^2/m=41.47$ MeV, $e^2=1.44$ MeV fm. The mass 
of the $\alpha$-particle is chosen as $M=3.973 m$, where
$m$ denotes the mass of the nucleon.
%%%%%%%%%%%%%%%%%%%%%%%%%%
The three body potential is taken to have Gaussian form
\begin{equation}
W(\rho) =V \exp\{-\rho^2/\beta^2 \},
\end{equation}
where $\rho^2 =\sum\limits_{i=1}^{3} {\bf r} _i^2 $,
$V=-31.935$ MeV and $\beta=3.315$ fm. Here
 ${\bf r}_i$ stands for the
position vector of $i$-th particle in the center of
mass frame of the three-$\alpha$ system.
%%%%%%%%%%%%%%%%%%%%%%%%%%%%
Since we consider here three identical particles we can reduce the
necessary Faddeev components to one.
We select states with total angular momentum
$L=0$. In Table 1 we show the convergence of the energy
of the ground state
and of the first resonant state with respect to $N$,
the number of CS functions employed in the expansion.
The selected resonance is the experimentally well-known
sharp state which has a great relevance in nuclear synthesis.

\subsection{The $H^-$ system}

In this system all the two-body interactions are of pure Coulomb type, two
of them are attractive, therefore we have to use the genuine Merkuriev approach.
Since the two electrons are identical particles we can reduce the number
of Faddeev components to two. Furthermore, we can define the cutoff function
such that  $v^{(s)}_{e  e}\equiv 0$.  In this case the
corresponding Faddeev component vanishes identically. So, finally
we have to deal with one Faddeev component only. 
In the $p  e$ subsystem we take the 
Merkuriev's cut with the functional form
\begin{equation}
\zeta(x,y) = 2/(1 + \exp((x/x_0)^\nu/(1+y/y_0))),
\end{equation}
and with the actual
parameters are $\nu=2.1$, $x_0=5$ and $y_0=10$. Fig.\ 2 and 3 show the
short and long range parts of $v_{pe}$, respectively.
In Table 2 we present the convergence of the energy
of a $L=0$ resonant state with respect to $N$ and the number
of angular momentum channels used in the bipolar expansion.

\section{Conclusions}
In this article we have presented a new method for calculating resonances
in three-body Coulombic systems. 
The homogeneous Faddeev-Merkuriev integral equations were solved for 
complex energies. For this, being an integral equation approach, 
no boundary conditions are needed. We solve the integral equations
by using the Coulomb-Sturmian separable expansion technique. 
The method works equally for
three-body systems with repulsive and attractive Coulomb interactions.

\section*{Acknowledgments}
This work has been supported by OTKA under Contracts No.\ T026233
and No.\ T029003 and by RFBR Grant No. 98-02-18190.

\begin{table}
\caption{Convergence of the ground-state and of the first resonant-state
 energy (in MeV) of a three-$\alpha$ system
interacting via the potential of (\ref{filpot})
with increasing basis for the separable expansion.
$N$ denotes the maximum number of
basis states employed for $n$ and $\nu$ in Eq. (\ref{sepfe}).
}
\label{tabc}
\begin{tabular}{rll}
\hline
$N$ & \multicolumn{1}{c}{\mbox{$E$} } &
\multicolumn{1}{c}{\mbox{$E=E_r-\mbox{i}\Gamma /2$} }  \\
\hline
16 & -7.283744 \phantom{000} & 0.3854244 -i 0.000011 \\
17 & -7.283779 & 0.3851242 -i 0.000011 \\
18 & -7.283801 & 0.3849323 -i 0.000012 \\
19 & -7.283815 & 0.3848056 -i 0.000012 \\
20 & -7.283824 & 0.3847236 -i 0.000012 \\
21 & -7.283829 & 0.3846683 -i 0.000012 \\
22 & -7.283833 & 0.3846308 -i 0.000012 \\
23 & -7.283836 & 0.3846053 -i 0.000012 \\
24 & -7.283837 & 0.3845873 -i 0.000013 \\
25 & -7.283838 & 0.3845748 -i 0.000013 \\
26 & -7.283839 & 0.3845658 -i 0.000013 \\
27 & -7.283840 & 0.3845593 -i 0.000013 \\
28 & -7.283840 & 0.3845546 -i 0.000013 \\
29 & -7.283640 & 0.3845512 -i 0.000013 \\
\hline
\end{tabular}
\end{table}

\begin{figure}
\psfig{file=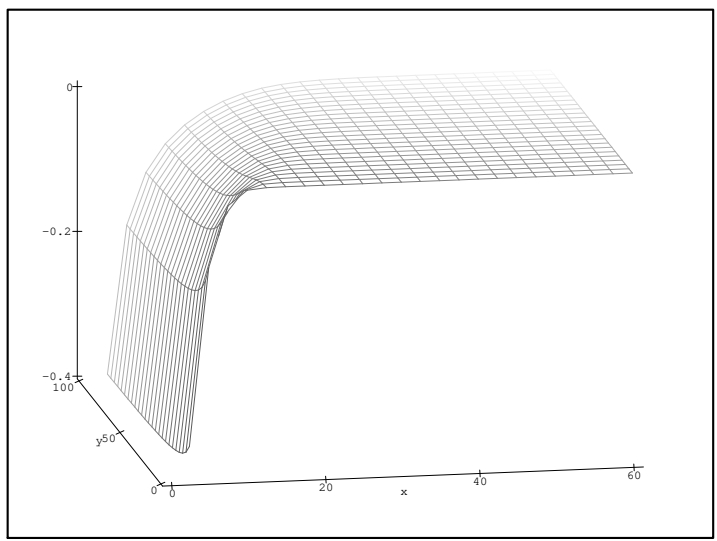,width=8cm,angle=-90}

\caption{The potential $v^{(s)}_{pe}$, the short-range
part of $v_{pe}$. }

\label{fig2}
\end{figure}

\begin{figure}
\psfig{file=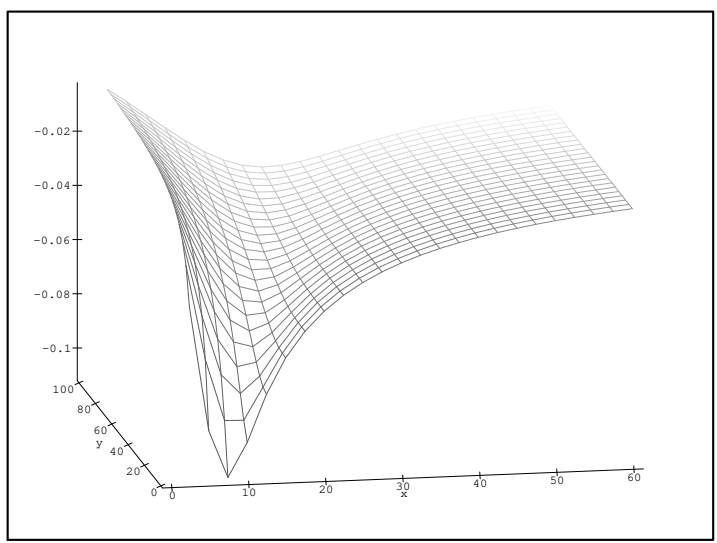,width=8cm,angle=-90}

\caption{The potential $v^{(l)}_{pe}$, the long-range
part of $v_{pe}$. }

\label{fig3}
\end{figure}

\begin{table}
\caption{Convergence  of a resonant state
of the $H^-$ system with respect to the number
of CS states. The angular momentum channels are taken into account
up to $l=2$, $l=3$ and $l=4$, respectively.
The (complex) energies are measured in atomic units.
}
\label{tabcc}
\begin{tabular}{rccc}
\hline
$N$ & $l=2$ & $l=3$ &  $l=4$   \\
\hline
 6 &  -0.146977  -i 0.000905 & -0.146989  -i 0.000903  & -0.146991  -i 0.000902  \\
 7 &  -0.147947  -i 0.000912 & -0.147957  -i 0.000910  & -0.147959  -i 0.000910  \\
 8 &  -0.148356  -i 0.000910 & -0.148365  -i 0.000908  & -0.148367  -i 0.000908  \\
 9 &  -0.148529  -i 0.000893 & -0.148538  -i 0.000892  & -0.148539  -i 0.000891  \\
10 &  -0.148608  -i 0.000882 & -0.148617  -i 0.000880  & -0.148618  -i 0.000880  \\
11 &  -0.148650  -i 0.000871 & -0.148659  -i 0.000870  & -0.148660  -i 0.000869  \\
12 &  -0.148669  -i 0.000871 & -0.148678  -i 0.000869  & -0.148680  -i 0.000869  \\
13 &  -0.148679  -i 0.000869 & -0.148688  -i 0.000868  & -0.148689  -i 0.000867  \\
14 &  -0.148683  -i 0.000870 & -0.148691  -i 0.000868  & -0.148693  -i 0.000868  \\
15 &  -0.148684  -i 0.000869 & -0.148693  -i 0.000867  & -0.148694  -i 0.000867  \\
16 &  -0.148685  -i 0.000869 & -0.148694  -i 0.000867  & -0.148695  -i 0.000867  \\
17 &  -0.148686  -i 0.000868 & -0.148695  -i 0.000867  & -0.148696  -i 0.000866  \\
18 &  -0.148686  -i 0.000869 & -0.148695  -i 0.000867  & -0.148696  -i 0.000866  \\
19 &  -0.148686  -i 0.000868 & -0.148695  -i 0.000867  & -0.148696  -i 0.000866  \\
20 &  -0.148686  -i 0.000868 & -0.148695  -i 0.000867  & -0.148696  -i 0.000866  \\

\hline
\end{tabular}
\end{table}

{}

\end{document}